\documentclass[12pt, a4paper]{article}
\usepackage{a4wide,amssymb,amsmath,amscd,graphicx,color}

\newcommand{\Z}{{\mathbb Z}}
\newcommand{\R}{{\mathbb R}}
\newcommand{\C}{{\mathbb C}}

\begin{document}

\topmargin -2pt

\headheight 0pt

\topskip 0mm \addtolength{\baselineskip}{0.20\baselineskip}
\begin{flushright}
{\tt KIAS-P10047}
\end{flushright}


\begin{center}
{\Large \bf Fermionic T-duality and Morita Equivalence  } \\

\vspace{5mm}

{\sc Ee Chang-Young}\footnote{cylee@sejong.ac.kr}\\
{\it Department of Physics, Sejong University, Seoul 143-747, Korea}\\

\vspace{3mm}

{\sc Hiroaki Nakajima}\footnote{nakajima@kias.re.kr}\\
{\it School of Physics, Korea Institute for Advanced Study,\\
207-43 Cheongnyangni-dong, Dongdaemun-gu,
Seoul 130-722, Korea,\\
and Department of Physics, Kyungpook National University,\\
Taegu 702-701, Korea} \\

\vspace{1mm}

and \\

\vspace{1mm}

{\sc Hyeonjoon Shin}\footnote{hyeonjoon@postech.ac.kr} \\
{\it Department of Physics,
   Pohang University of Science and Technology, \\
   and Asia Pacific Center for Theoretical
  Physics, \\
  Pohang, Gyeongbuk 790-784, Korea} \\

\vspace{5mm}


{\bf ABSTRACT} \\
\end{center}


\noindent
 In this paper we investigate the relationship between the so-called fermionic
 T-duality and the
 Morita equivalence of noncommutative supertori.
 We first get an action satisfying the BRST invariance under nonvanishing
 constant R-R and NS-NS backgrounds in the hybrid formalism.
 We investigate the effect of bosonic T-duality transformation together with
 fermionic T-duality
 transformation in this background
 and look for the resultant symmetry of transformations.
We find that the duality transformations correspond to
Morita equivalence of noncommutative supertori.
In particular, we obtain the symmetry group $SO(2,2,{\cal V}_{\Z}^0)$
in two dimensions,
where ${\cal V}_{\Z}^0$ denotes Grassmann even number
whose body part belongs to ${\Z}$. 
  \\

\setcounter{footnote}{0}

 \vfill

\noindent

\thispagestyle{empty}

\newpage
\section{Introduction}

One typical symmetry of string theory is T-duality.  It originates from
the symmetry of string worldsheet action under a shift of string
coordinate and relates string theories on generically different target
space-time backgrounds.  At the level of string spectrum, T-duality
exchanges the momentum and the winding mode. Recently, besides this
usual T-duality which is now referred as the bosonic T-duality, a different
kind of T-duality has been proposed in the process of understanding the
dual superconformal symmetry displayed by planar scattering amplitudes of
super Yang-Mills theory from the viewpoint of superstring theory
\cite{bm08,ft2}.  Its origin is the symmetry of tree level superstring
theory under a shift of fermionic coordinate rather than bosonic one, and
thus it is dubbed fermionic T-duality.  Similar to the bosonic T-duality,
the fermionic T-duality maps superstring theory on one supersymmetric
background to that on another supersymmetric background.

As a new kind of duality, one may be curious about the fermionic
T-duality itself.  Indeed, there have been works exploring  properties
and various aspects of the fermionic T-duality as follows.
It has been shown that some basic symmetry structures of pure spinor
string theory are preserved under the fermionic T-duality
transformation up to one-loop level \cite{Chandia:2009yv}.
The self-duality of the supercoset sigma model description
of superstring has been investigated in \cite{Adam:2009kt}.
In the context of supergravity, the problem of complexification of
supergravity fields after the fermionic T-duality has been
considered in \cite{Bakhmatov:2009be}. Supersymmetric generalization
of duality, superduality, which may connect to the fermionic T-duality
has been also given \cite{Fre:2009ki}.

As alluded to above, the fermionic T-duality is quite similar to the bosonic
one from the worldsheet viewpoint.  Actually, this is also the case in the
canonical formulation of the fermionic T-duality \cite{Sfetsos:2010xa},
where the duality transformation is formally represented as the exchange of
momentum and winding.  If we now recall the well established fact that the
bosonic T-duality is related to the Morita equivalence of noncommutative tori
\cite{rfsz98,schwarz98},
such similarity opens up the possiblity of uncovering the mathematical
structure of the fermionic T-duality via Morita equivalence.
We note that in the bosonic noncommutative torus case the metric is flat and the
 background NS-NS B-field is constant.
 Incidentally, the  NS-NS B-field is related with noncommutativity
of space and with the T-duality \cite{cds,SeWi}, while the graviphoton flux is
related with nonanticommutativity of  space \cite{Se,BeSe} and with the
  fermionic T-duality \cite{bm08}.
In this paper, we will investigate underlying mathematical structure of the fermionic
T-duality with a flat metric,  a constant B-field, and a constant self-dual graviphoton flux.

Now, let us give a brief historical description of the bosonic T-duality
in connection with noncommutativity.
 Connes, Douglas, and Schwarz \cite{cds} first showed that
  two dimensional noncommutative tori can emerge
 from toroidal compactifications of M(atrix) theory
with nonvanishing NS-NS field backgrounds.
Then, Rieffel and Schwarz \cite{rfsz98} showed that the actions of the group
$SO(n,n,{\Z})$ on an antisymmetric $n\times n$ matrix  which
represents noncommutativity parameters for an $n$-dimensional noncommutative
torus yield Morita equivalent tori. A subsequent work by Schwarz
\cite{schwarz98} showed that compactifications on Morita equivalent
tori are physically equivalent, corresponding to T-duality in string theory.
This bosonic T-duality usually connects different NS-NS field backgrounds.
However note that, as was shown in the Green-Schwarz formalism
\cite{Cvetic:1999zs} as well as in the pure spinor formalism
\cite{Benichou:2008it},
even the bosonic T-duality transformations can connect different R-R field backgrounds.\footnote{We would like to note that the bosonic T-duality
changes the form degree of the R-R fields in the usual case
while the fermionic T-duality
shifts the value of the R-R fields.}
Therefore a decisive factor for the bosonic or fermionic T-duality is not the background fields
but the symmetry transformations via which dual theories are connected:
in the bosonic T-duality case, it is a shift symmetry along  bosonic coordinates,
and in the fermionic T-duality case, it is a shift symmetry along  fermionic coordinates.

Since the bosonic T-duality is related with the Morita equivalence of noncommutative tori
representing different NS-NS backgrounds,
we wonder whether the fermionic T-duality relating different R-R field backgrounds
\cite{bm08} is related with the Morita equivalence of noncommutative supertori \cite{ekn10}.
 For this purpose, we investigate the fermionic T-duality transformations under the presence of
 both R-R and NS-NS background fields.
 Then we compare the obtained symmetry of the above T-duality transformations with the symmetry
 between noncommutatve supertori \cite{ekn08} related by Morita equivalence.
 In a flat geometry with a constant B-field, and a constant self-dual graviphoton flux, we show that
 the fermionic T-duality corresponds to the Morita equivalence of noncommutative supertori
 in two and four dimensions.

This paper is organized as follows.
In section 2, we investigate the fermionic T-duality transformations in the presence of NS-NS
and R-R background fields. In section 3, we look into the relationship between this duality symmetry
and the Morita equivalence of noncommutative supertori.
We conclude with discussion in section 4.
\\

\section{Bosonic and Fermionic T-duality in NS-NS and R-R Backgrounds}

We start with type II superstring compactified on Calabi-Yau three-fold,
where the background of the constant NS-NS B-field $B_{\mu\nu}$
and the constant self-dual graviphoton field strength $F^{\alpha\beta}$
are turned on in four-dimensional spacetime.
We consider the case such that the closed string metric $g_{\mu\nu}$ and
the dilaton $\phi$ are also constant. Then we neglect the dilaton coupling
$\sim\int d^{2}z \phi R^{(2)}$.
The worldsheet action can be explicitly written down
using the hybrid formalism \cite{Be}
(or pure spinor formalism in four dimensions \cite{Wi}) as
\begin{align}
S
&=
\frac{1}{2\pi\alpha'}\int d^{2}z\biggl[\frac{1}{2}
(g_{\mu\nu}+2\pi\alpha' B_{\mu\nu})\partial X^{\mu}\tilde{\partial}X^{\nu}
\notag\\
&\qquad\qquad\qquad\quad{}
+p_{\alpha}\tilde{\partial}\theta^{\alpha}
+\bar{p}_{\dot{\alpha}}\tilde{\partial}\bar{\theta}^{\dot{\alpha}}
+\tilde{p}_{\alpha}\partial\tilde{\theta}^{\alpha}
+\tilde{\bar{p}}_{\dot{\alpha}}\partial\tilde{\bar{\theta}}^{\dot{\alpha}}
+2\pi\alpha' F^{\alpha\beta}q_{\alpha}\tilde{q}_{\beta}
\biggr]+S^{}_{C},
\label{ws1}
\end{align}
where $\mu=0,\ldots,3$ and $\alpha,\dot{\alpha}=1,2$.
$p$ denotes the conjugate momentum of fermionic coordinate $\theta$.
We use the tilde to express the worldsheet chirality, while the bar is used
to express the chirality in four dimensions.
$S^{}_{C}$ consists of the action for the chiral bosons \cite{Be}
(or the action for the pure spinor in the pure spinor formalism \cite{Wi})
and the action for the compactified direction.
$q_{\alpha}$, $\tilde{q}_{\alpha}$ are the chiral supercharges as
worldsheet variables, defined by
\begin{align}
q_{\alpha}
&=-p_{\alpha}-i\sigma^{\mu}_{\alpha\dot{\alpha}}\bar{\theta}^{\dot{\alpha}}
\partial X_{\mu}+\frac{1}{2}\bar{\theta}\bar{\theta}\partial\theta_{\alpha}
-\frac{3}{2}\partial(\theta_{\alpha}\bar{\theta}\bar{\theta}),
\notag\\
\tilde{q}_{\alpha}
&=-\tilde{p}_{\alpha}-i\sigma^{\mu}_{\alpha\dot{\alpha}}
\tilde{\bar{\theta}}^{\dot{\alpha}}\tilde{\partial}X_{\mu}
+\frac{1}{2}\tilde{\bar{\theta}}\tilde{\bar{\theta}}
\tilde{\partial}\tilde{\theta}_{\alpha}
-\frac{3}{2}\tilde{\partial}
(\tilde{\theta}_{\alpha}\tilde{\bar{\theta}}\tilde{\bar{\theta}}).
\label{scws}
\end{align}
Note that in the case of $B_{\mu\nu}=0$, the action \eqref{ws1} is reduced to
the one discussed in \cite{OoVa, Se, BeSe}.


We introduce the chiral coordinate $Y^{\mu}$ as
\begin{gather}
Y^{\mu}=X^{\mu}
+i\theta^{\alpha}\sigma^{\mu}_{\alpha\dot{\alpha}}\bar{\theta}^{\dot{\alpha}}
+i\tilde{\theta}^{\alpha}\sigma^{\mu}_{\alpha\dot{\alpha}}
\tilde{\bar{\theta}}^{\dot{\alpha}}.
\label{chiral}
\end{gather}
The worldsheet action \eqref{ws1} can be rewritten in terms of $Y^{\mu}$ as
\begin{align}
S
&=
\frac{1}{2\pi\alpha'}\int d^{2}z\biggl[\frac{1}{2}
(g_{\mu\nu}+2\pi\alpha' B_{\mu\nu})\partial Y^{\mu}\tilde{\partial}Y^{\nu}
\notag\\
&\qquad\qquad\quad{}
-q_{\alpha}\tilde{\partial}\theta^{\alpha}
+\bar{d}_{\dot{\alpha}}\tilde{\partial}\bar{\theta}^{\dot{\alpha}}
-\tilde{q}_{\alpha}\partial\tilde{\theta}^{\alpha}
+\tilde{\bar{d}}_{\dot{\alpha}}\partial\tilde{\bar{\theta}}^{\dot{\alpha}}
+2\pi\alpha' F^{\alpha\beta}q_{\alpha}\tilde{q}_{\beta}
\biggr]+\text{(surface term)},
\label{ws2}
\end{align}
where $\bar{d}_{\dot{\alpha}}$, $\tilde{\bar{d}}_{\dot{\alpha}}$ are the
anti-chiral supercovariant derivatives as worldsheet variables,
defined by
\begin{align}
\bar{d}_{\dot{\alpha}}
&=\bar{p}_{\dot{\alpha}}-i\theta^{\alpha}\sigma^{\mu}_{\alpha\dot{\alpha}}
\partial X_{\mu}-\theta\theta\partial\bar{\theta}_{\dot{\alpha}}
+\frac{1}{2}\bar{\theta}_{\dot{\alpha}}\partial(\theta\theta),
\notag\\
\tilde{\bar{d}}_{\dot{\alpha}}
&=\tilde{\bar{p}}_{\dot{\alpha}}
-i\tilde{\theta}^{\alpha}\sigma^{\mu}_{\alpha\dot{\alpha}}
\tilde{\partial} X_{\mu}
-\tilde{\theta}\tilde{\theta}\partial\tilde{\bar{\theta}}_{\dot{\alpha}}
+\frac{1}{2}\tilde{\bar{\theta}}_{\dot{\alpha}}\tilde{\partial}
(\tilde{\theta}\tilde{\theta}).
\end{align}
In the action \eqref{ws2},
$q_{\alpha}$, $\tilde{q}_{\alpha}$ and
$\bar{d}_{\dot{\alpha}}$, $\tilde{\bar{d}}_{\dot{\alpha}}$ are regarded
as the conjugate momenta of $\theta$'s and are independent of $Y^{\mu}$.
Then the action is now quadratic in terms of these variables
and there is no backreaction.
This consistency can be also explained if we assume that
the graviphoton field strength
$F^{\alpha\beta}$ comes from the R-R five-form field strength
in ten-dimensional type IIB superstring.
This assumption is consistent
with the fact that supersymmetric Yang-Mills theory on
non(anti)commutative $\mathcal{N}=1$ superspace,
in which non(anti)commutativity is induced by $F^{\alpha\beta}$,
is reproduced by the coupling between D-branes and
the R-R five-form field strength \cite{Ito:2006tw}.
This coupling can be determined through
the calculation of the worldsheet disk amplitudes.
Under the assumption, it is shown that the backgrounds satisfy the equation
of motion.
On the other hand, other R-R backgrounds induce the different kind of
deformation,
such as $\Omega$-background deformation from R-R three-form field strength
\cite{omega}.

We consider the bosonic and fermionic T-duality on this system.
Here we restrict ourselves to the case that the worldsheet does not have any
nontrivial cycle because the fermionic T-duality may not be a symmetry for
the worldsheet with nonzero genus as discussed in \cite{bm08}.
Since in the worldsheet action \eqref{ws2}
the bosonic part (first line) and the fermionic part (second line)
are decoupled, we can apply the Buscher procedure \cite{Bu}
to two parts separately.
For bosonic part, we decompose $Y^{\mu}=(\hat{Y}^{i}, \check{Y}^{l})$
and we apply the duality transformation
to $\hat{Y}^{i}$. We also decompose the NS-NS background
$E_{\mu\nu}=\frac{1}{2\pi\alpha'}g_{\mu\nu}+B_{\mu\nu}$ as
\begin{gather}
E_{\mu\nu}=
\begin{pmatrix}
E_{a} & E_{b} \\
E_{c} & E_{d}
\end{pmatrix}.
\label{part-dual}
\end{gather}
Since $E_{\mu\nu}$ is constant, we have a shift isometry
$\hat{Y}^{i}\to\hat{Y}^{i}+y^{i}$.
We rewrite the bosonic part of the worldsheet action \eqref{ws2}
by introducing the gauge fields
$A$ and $\tilde{A}$
with the constraint $\tilde{\partial}A-\partial\tilde{A}=0$:
\begin{align}
S_{B}
&=
\frac{1}{2}\int d^{2}z\,
\Bigl[
A^{t}E_{a}\tilde{A}
+A^{t}E_{b}\tilde{\partial}\check{Y}
+\partial\check{Y}^{t}E_{c}\tilde{A}
+\partial\check{Y}^{t}E_{d}\tilde{\partial}\check{Y}
+\hat{Y}^{\prime t}\bigl(\tilde{\partial}A-\partial\tilde{A}\bigr)
\Bigr].
\end{align}
By integrating out $\hat{Y}'$, we have
$A=\partial\hat{Y}$ and $\tilde{A}=\tilde{\partial}\hat{Y}$
for some $\hat{Y}$ and the above action comes back to the original form.
Instead,
by reshuffling the terms,
one can show that
\begin{align}
S_{B}
&=
\frac{1}{2}\int d^{2}z\,
\Bigl[
\bigl(A^{t}+(\partial\check{Y}E_{c}^{}+\partial Y^{\prime t})E_{a}^{-1}\bigr)
E_{a}^{}
\bigl(\tilde{A}+E_{a}^{-1}(E_{b}^{}\tilde{\partial}\check{Y}
-\tilde{\partial}Y')\bigr)
\notag\\
&\quad{}
+\partial\hat{Y}^{\prime t}E_{a}^{-1}\tilde{\partial}\hat{Y}'
-\partial\hat{Y}^{\prime t}E_{a}^{-1}E_{b}^{}\tilde{\partial}\check{Y}
+\partial\check{Y}^{t}E_{c}^{}E_{a}^{-1}\tilde{\partial}\hat{Y}'
+\partial\check{Y}^{t}(E_{d}^{}-E_{c}^{}E_{a}^{-1}E_{b}^{})
\tilde{\partial}\check{Y}
\Bigr],
\label{ws3}
\end{align}
where the first line in \eqref{ws3} is integrated out
and it generates the shift of dilaton $\phi$ as
$\phi\to\phi-\frac{1}{2}\log\det E_{a}$. Then the dilaton is
again constant. From the second line one can read off the transformation
of $E_{\mu\nu}$ as
\begin{gather}
E_{\mu\nu}\to E'_{\mu\nu}=
\begin{pmatrix}
E_{a}^{-1}      & -E_{a}^{-1}E_{b}^{}           \\
E_{c}^{}E_{a}^{-1} & E_{d}^{}-E_{c}^{}E_{a}^{-1}E_{b}^{}
\end{pmatrix}.
\label{Etrans}
\end{gather}
Note that the T-duality transformation which we examined above is slightly
different from the conventional one since we take the transformation for
the chiral coordinate $Y^{\mu}$, while we usually consider the transformation
for $X^{\mu}$. The transformation of $E_{\mu\nu}$ is the same as
the conventional one but $F^{\alpha\beta}$ is unchanged%
\footnote{
In conventional T-duality in terms of $X^{\mu}$,
the rank of R-R fields is changed due to the flipping of the spinor chirality.
}
in our case
because of the decoupling between the bosonic part and the fermionic part
in the action \eqref{ws2}.

For fermionic part, we apply the fermionic T-duality transformation
given in \cite{bm08}. Although we have the graviphoton background
$F^{\alpha\beta}$, it turns out that one can perform the transformation
in the same way as in \cite{bm08} because the term containing $F^{\alpha\beta}$
depends only on $q_{\alpha}$ and $\tilde{q}_{\alpha}$.
Since the action \eqref{ws2} does not contain the square of the derivative,
we add the following surface term to the action \eqref{ws2}
\begin{gather}
S_{\mathrm{b}}=
\frac{1}{(2\pi\alpha')^{2}}\int d^{2}z (f^{-1})_{\alpha\beta}\bigl(
\partial\theta^{\alpha}\tilde{\partial}\tilde{\theta}^{\beta}
-\tilde{\partial}\theta^{\alpha}\partial\tilde{\theta}^{\beta}
\bigr),
\label{st}
\end{gather}
where $f^{\alpha\beta}=f^{\beta\alpha}$ is constant.
Since the background preserves chiral supersymmetry
which can be regarded as a shift isometry in the fermionic
direction $\theta^{\alpha}\to\theta^{\alpha}+\rho^{\alpha}$,
$\tilde{\theta}^{\alpha}\to\tilde{\theta}^{\alpha}+\tilde{\rho}^{\alpha}$,
then one can dualize
$\theta^{\alpha}$ and $\tilde{\theta}^{\alpha}$ by introducing
the fermionic gauge fields
$(\mathcal{A}^{\alpha},\tilde{\mathcal{A}}^{\alpha})$ and
$(\hat{\mathcal{A}}^{\alpha},\hat{\tilde{\mathcal{A}}}^{\alpha})$
with the constraints
$\tilde{\partial}\mathcal{A}^{\alpha}-\partial\tilde{\mathcal{A}}^{\alpha}=
\tilde{\partial}\hat{\mathcal{A}}^{\alpha}
-\partial\hat{\tilde{\mathcal{A}}}^{\alpha}=0$
as
\begin{align}
S_{F}
&=
\frac{1}{2\pi\alpha'}\int d^{2}z \Bigl[
-q_{\alpha}\tilde{\mathcal{A}}^{\alpha}
-\tilde{q}_{\alpha}\hat{\mathcal{A}}^{\alpha}
+(2\pi\alpha')^{-1}(f^{-1})_{\alpha\beta}\bigl(
\mathcal{A}^{\alpha}\tilde{\hat{\mathcal{A}}}^{\beta}
-\tilde{\mathcal{A}}^{\alpha}\hat{\mathcal{A}}^{\beta})
+2\pi\alpha'F^{\alpha\beta}q_{\alpha}\tilde{q}_{\beta}
\notag\\
&\qquad\qquad\qquad\quad{}
+\chi_{\alpha}
(\tilde{\partial}\mathcal{A}^{\alpha}-\partial\tilde{\mathcal{A}}^{\alpha})
+\hat{\chi}_{\alpha}
(\tilde{\partial}\hat{\mathcal{A}}^{\alpha}
-\partial\hat{\tilde{\mathcal{A}}}^{\alpha})
+\bar{d}_{\dot{\alpha}}\tilde{\partial}\bar{\theta}^{\dot{\alpha}}
+\tilde{\bar{d}}_{\dot{\alpha}}\partial\tilde{\bar{\theta}}^{\dot{\alpha}}
\Bigr],
\label{ws4}
\end{align}
where $\chi_{\alpha}$ and $\hat{\chi}_{\alpha}$ are
the fermionic Lagrange multipliers for the above constraints.
By integrating out the fermionic gauge fields,
the action \eqref{ws4} becomes
\begin{align}
S_{F}
&=
\frac{1}{2\pi\alpha'}\int d^{2}z \Bigl[
-2\pi\alpha'f^{\alpha\beta}
\bigl(\partial\hat{\chi}_{\alpha}\tilde{\partial}\chi_{\beta}
-\tilde{\partial}\hat{\chi}_{\alpha}\partial\chi_{\beta}\bigr)
+2\pi\alpha'\tilde{q}_{\alpha}f^{\alpha\beta}\partial\chi_{\beta}
+2\pi\alpha'q_{\alpha}f^{\alpha\beta}\tilde{\partial}\hat{\chi}_{\beta}
\notag\\
&\qquad\qquad\qquad\quad{}
+2\pi\alpha'(F^{\alpha\beta}+f^{\alpha\beta})q_{\alpha}\tilde{q}_{\beta}
+\bar{d}_{\dot{\alpha}}\tilde{\partial}\bar{\theta}^{\dot{\alpha}}
+\tilde{\bar{d}}_{\dot{\alpha}}\partial\tilde{\bar{\theta}}^{\dot{\alpha}}
\Bigr].
\label{dual}
\end{align}
The first term is the surface term and we drop it out.
The dual fermionic coordinates are identified as
\begin{gather}
\theta^{\prime\alpha}=-2\pi\alpha'f^{\alpha\beta}\chi_{\beta},\quad
\tilde{\theta}^{\prime\alpha}=-2\pi\alpha'f^{\alpha\beta}\hat{\chi}_{\beta}.
\end{gather}
%
$F^{\alpha\beta}$ is just shifted by the constant $f^{\alpha\beta}$ as
\begin{gather}
F^{\alpha\beta}\to F^{\prime\alpha\beta}=F^{\alpha\beta}+f^{\alpha\beta}.
\label{Ftrans}
\end{gather}
We also have the constant shift of the dilaton from the integration
of the fermionic gauge fields.

In the next section, we will show that the bosonic and fermionic
T-duality transformations \eqref{Etrans}, \eqref{Ftrans}
and other symmetries are combined into the supersymmetric version of
Morita equivalence. 
In the case of $F^{\alpha\beta}=0$,
when we put the D-branes filling the four-dimensional spacetime,
the backgrounds $B_{\mu\nu}$ induces
noncommutativity on the D-branes
\cite{Chu:1998qz, SeWi}.
Under the limit
\begin{gather}
g_{\mu\nu}\sim(\alpha')^{2},\quad B_{\mu\nu}\text{: finite} \quad
\mbox{for} \quad \alpha' \to 0,
\label{SWlimit}
\end{gather}
the non(anti)commutativity among the coordinates are
\begin{gather}
[X^{\mu}, X^{\nu}]=\Theta^{\mu\nu} \quad \text{or} \quad
[Y^{\mu}, Y^{\nu}]=\Theta^{\mu\nu}.
\label{ncparameters}
\end{gather}
Here the noncommutativity parameters $\Theta^{\mu\nu}$ ($\Theta^t = -\Theta$)
and the open string metric $G_{\mu\nu}$
are respectively obtained as
\cite{Chu:1998qz, SeWi}
\begin{gather}
\Theta^{\mu\nu}=(B^{-1})^{\mu\nu},\quad
G_{\mu\nu}=-(2\pi\alpha')^{2}(Bg^{-1}B)_{\mu\nu}.
\label{NC}
\end{gather}
Then the duality transformations of the parameters $\Theta^{\mu\nu}$
can be obtained from \eqref{Etrans}
as
\begin{gather}
\Theta^{\mu\nu}=
\begin{pmatrix}
\Theta_{a} & \Theta_{b} \\
\Theta_{c} & \Theta_{d}
\end{pmatrix}
\to
\Theta^{\prime\mu\nu}=
\begin{pmatrix}
\Theta_{a}^{-1}      & -\Theta_{a}^{-1}\Theta_{b}^{}           \\
\Theta_{c}^{}\Theta_{a}^{-1} & \Theta_{d}^{}
-\Theta_{c}^{}\Theta_{a}^{-1}\Theta_{b}^{}
\end{pmatrix}
\label{Etrans2}
\end{gather}
The above and other symmetries are combined into $SO(n,n,\mathbb{Z})$
group which gives the Morita equivalence of noncommutative tori
\cite{rfsz98,schwarz98}.

When $F^{\alpha\beta}$ is also turned on,
the fermionic coordinates become nonanticommutative
\cite{OoVa, Se, BeSe} as
\begin{gather}
[Y^{\mu}, Y^{\nu}]=\Theta^{\mu\nu}.
\quad
\{\theta^{\alpha},\theta^{\beta}\}=C^{\alpha\beta}, \quad
\text{(others)}
=0,
\label{ncparameters2}
\end{gather}
where the nonanticommutativity parameter $C^{\alpha\beta}$ is defined by
\begin{gather}
C^{\alpha\beta}=(2\pi\alpha')^{2}F^{\alpha\beta}.
\end{gather}
Then if we assume that we can apply the result of the duality transformation
\eqref{Ftrans}, $C^{\alpha\beta}$ transforms as
\begin{gather}
C^{\alpha\beta}\to C^{\prime\alpha\beta}=C^{\alpha\beta}+c^{\alpha\beta},
\label{Ftrans2}
\end{gather}
where $c^{\alpha\beta}=(2\pi\alpha')^{2}f^{\alpha\beta}$.
Here we note that
under the fermionic T-duality transformation
adding surface term \eqref{st} and removing the first term from
\eqref{dual} change the boundary condition  of the D-branes (see appendix).
However, if we start from the full ten-dimensional pure spinor formalism,
we do not need to add or remove the surface term since the square of
the derivative of $\theta$ is already present, and thus
the problem with the boundary condition does not occur.
It turns out that in this case the R-R background is
shifted by the square of the constant Killing spinor%
\footnote{In our case, the Killing spinor is constant
in ten-dimensional pure spinor formalism. Then the shift is also constant.}
\cite{bm08}.
So we use the transformation rule \eqref{Ftrans2} even under the presence
of the D-branes.


\section{Relation to Morita Equivalence} \label{Morita-super}

Before considering the relationship between the fermionic T-duality
and Morita equivalence, we first briefly review the Morita
equivalence of noncommutative supertori \cite{ekn10}.
An $n$-dimensional noncommutative torus($A_\theta^n$) is an associative algebra with involution having
unitary generators $U_1, \dots , U_n$ obeying the relations
\begin{align}
U_iU_j=e^{2 \pi i \theta_{ij}}U_jU_i , ~~~ i,j=1,\dots,n,
\label{nctorus}
\end{align}
where the noncommutativity parameters $\theta_{ij}$ form a real $n\times n$ anti-symmetric matrix $\Theta$.
%


The
endomorphism algebra of the module of noncommutative torus
 is Morita equivalent to the given noncommutative torus.
When $U_i$'s belong to the given noncommutative torus
and $Z_i$'s belong to its endomorphism algebra,
then $Z_i$'s commute with the $U_i$'s, i.e.,
\begin{equation}
 U_i Z_j = Z_j U_i   \; \; {\rm where} \; \; Z_i Z_j=e^{2 \pi i \theta_{ij}'}Z_j Z_i , ~~~ i,j=1,\dots,n .
 \label{dualtrs}
\end{equation}
Let $D$ be a lattice in ${\cal G}=M\times \hat{M}$, where
$M={\R}^p\times{\Z}^q$ with $2p+q=n$ and $\hat{M}$ is its dual.
The embedding map $\Phi$ under which $D$ is the image of ${\Z}^n$
 determines a projective module $E$ on which the algebra of noncommutative torus acts.
In the Heisenberg representation the operators $U_i$ can be
 defined by
\begin{equation}
U_{(m,\hat s)}f(r)=e^{2\pi i <r, \hat s>}f(r+m), ~~ m,r \in M, ~ \hat s \in \hat{M}, ~
  f \in E ,
\label{heisrep}
\end{equation}
where we denoted $U_i := U_{\vec{e}_i}$ with $\vec{e}_i :=(m,\hat s)$,
and $<r, \hat s>$ is usual inner product between $M$ and $\hat{M}$.
%
%
In this representation, we  get
\begin{eqnarray}
U_{(m,\hat s)}U_{(n,\hat t)} =   e^{2\pi i (<m, \hat t> - <n, \hat s>)}U_{(n,\hat t)}U_{(m,\hat s)} .
\label{cocycler}
\end{eqnarray}
Denoting  the basis as $\vec{e}_i :=(m,\hat s), \; \vec{e}_j :=(n,\hat t)$ and
the embedding map as
$ \Phi =( \vec{e}_1, \vec{e}_2, \cdots, \vec{e}_n )$, then
 $\theta_{ij} $ can be expressed as\footnote{Here we assume $n$ is even and $n=2p$.}
 \begin{align}
\theta_{ij} = \vec{e}_i \cdot  J_0  \vec{e}_j  \ \ {\rm where } \  \ J_0 =\begin{pmatrix} 0 & I_p \\ -I_p & 0 \end{pmatrix} ,
\label{cyctheta}
\end{align}
and thus \cite{ekn10,ekn08}
 \begin{align}
\Theta = \Phi^t  J_0  \Phi  .
\label{Theta}
\end{align}
%
%
Therefore one can see that the condition for  Morita equivalent dual torus \eqref{dualtrs}
can be written as \cite{rfsz98,ekn10}
\begin{equation}
 \Phi^t J_0 \Phi' = K ,
\label{b_dual}
\end{equation}
where $\Phi$ is the embedding map of a given torus and $\Phi'$
is the embedding map of the dual torus,
and $K$ is an $n \times n$
matrix whose elements belong to ${\Z}$.

In the supersymmetric case, one can put the relation \eqref{b_dual}
into the following form \cite{ekn10}:
\begin{equation}
\widetilde{\Phi}^t \widetilde{J}_0 \widetilde{\Phi}' = B^t J_0  B' + F^t \hat{J}_0 F' = \widetilde{K},
 \ \ {\rm where} \
\hat{J}_0 =
\begin{pmatrix} 0 & I_r \\ I_r & 0 \end{pmatrix}.
\label{s_dual}
\end{equation}
Here
 $\widetilde{\Phi}:=\begin{pmatrix} B \\ F \end{pmatrix}$ and
 $\widetilde{\Phi}':=\begin{pmatrix} B' \\ F' \end{pmatrix}$ are
 the embedding maps of the given supertorus and the dual supertorus respectively,
 the elements of the matrix  $\widetilde{K}$ belong to $ {\Z}$, and $r$ depends on
 the number of supersymmetry generators \cite{ekn08}.
The entries of  $B, B'$ and $F, F'$ belong to ${\cal V}^0$ and
${\cal V}^1$ respectively, which are the even and odd parts of a Grassmann
algebra  ${\cal V}$ over ${\C}$.
In the supersymmetric case, we denote the
noncommutativity parameter matrix as $\widetilde{\Theta}$
whose elements
 belong to ${\cal V}^0$:
\begin{equation}
\widetilde{\Theta}= \widetilde{\Phi}^t \widetilde{J}_0 \widetilde{\Phi} = B^t J_0 B + F^t \hat{J}_0 F.
\label{theta_BF}
\end{equation}
%
%
%
%
Note that under the change of basis,
the matrix $\widetilde{K}$ in the duality condition \eqref{s_dual}
 can be any element in $GL(n,{\Z})$.\footnote{$\Phi$ and  $\Phi'$ in Eq. (28)
 can be regarded as different sets of basis vectors which yield commuting generators
 $U_i$'s and $Z_i$'s. Thus the inner product between the two bases which have the same rank
should be nonsingular. Therefore together with the condition for a dual torus, $K$ should be a nonsingular
$n \times n $ integer matrix. The same holds for  $\tilde{ \Phi}$ and $\tilde{ \Phi}'$ with $\tilde{K}$.}
 Below we will denote the body and soul parts of $\widetilde{\Theta}$ as
  $ B^t J_0 B := \Theta_B $  and $ F^t \hat{J}_0 F := \Theta_F $.

 From the condition  \eqref{s_dual} we can express the bosonic part of the dual embedding map as
\begin{equation}
B'= - J_0 B^{-t}(\widetilde{K}-F^t \hat{J}_0 F').
\label{dual_b}
\end{equation}
 Using the relations  \eqref{theta_BF} and \eqref{dual_b},
 we can express the noncommutativity parameter matrix $\widetilde{\Theta}' $ of the dual supertorus as
\begin{eqnarray}
\widetilde{\Theta}'
&=& -(\widetilde{K}-F^t \hat{J}_0 F')^t (B^t J_0 B)^{-1}
(\widetilde{K}-F^t \hat{J}_0 F')
+F^{'t} \hat{J}_0 F',
\label{s_btheta}
\end{eqnarray}
where we used
 $J_0^{-1} = - J_0$.
This shows that the duality condition \eqref{s_dual} does not restrict
 the  soul part $F'$ of the dual map $\widetilde{\Phi}'$ in  the defining relation \eqref{s_btheta}
  for the noncommutativity parameters $\widetilde{\Theta}' $
 of the dual torus.
Therefore, as it was stressed in \cite{ekn10},
when two $\widetilde{\Theta}$'s have the same body parts and only differ over the soul parts,
then the two corresponding tori are Morita equivalent.

At this point, we want to consider the result obtained by the T-dual transformations in the previous
section.
There the noncommutativity parameters were given by \eqref{ncparameters}.
In the present notation $\Theta^{\mu\nu}, C^{\alpha\beta}$ together give
 $\widetilde{\Theta}=\Theta \oplus  C $. Namely, $ \Theta$ in the previous section corresponds
 to $\Theta_B$ here, and $C$ corresponds to $\Theta_F$.
 If we only consider the T-dual transformation of $\hat{Y}^{i}$ coordinates  in the previous section,
 then we end up with $\Theta \rightarrow \Theta^{-1} $ as it was shown in \eqref{Etrans2}.
 This agrees with \eqref{s_btheta}, which is what we get when both $F$ and $F'$ vanish,
 up to an allowed $GL(n,{\Z})$ transformation of $\widetilde{K}$ as we mentioned above.
 For the fermionic part,  $C $ changes by a shift in  \eqref{Ftrans2}, which agrees
 with our above statement that the two tori with the same body parts are Morita equivalent.\footnote{
 This conforms with the result in section 2, Eq.\eqref{Ftrans}, that $F^{\alpha\beta}$ can be shifted
 by an arbitrary constant $f^{\alpha\beta}$. Such shift does not affect the physics at the string
 tree-level \cite{bm08}.}
 Therefore  the T-dual, both bosonic and fermionic, transformations correspond to
 the Morita equivalence of  noncommutative supertori represented by $\Theta$'s  and  $C$'s.

In order to analyze the symmetry structure of the above Morita equivalence
we first consider the following linear fractional transformation, an action of
an element of $SO(n,n, {\cal V}_{\Z}^0) $
where ${\cal V}_{\Z}^0$ denotes Grassmann even number whose body part belongs to ${\Z}$.
\begin{equation}
 g \widetilde{\Theta}:= (A \widetilde{\Theta} + B)(C \widetilde{\Theta} +D)^{-1} \ \ \
 {\rm with} \ \
 g =\begin{pmatrix} A& B\cr  C&D
  \end{pmatrix} \in SO(n,n, {\cal V}_{\Z}^0).
 \label{gaction_theta}
\end{equation}
Now we consider a case when $g$ becomes
$\sigma_n=\begin{pmatrix} 0 & I_n  \\ I_n &
0 \end{pmatrix} $ with $n$ denoting the dimension of a torus:
\begin{equation}
 \sigma_n \widetilde{\Theta} =  \widetilde{\Theta}^{-1} = ( \Theta_B + \Theta_F )^{-1}
 = \widetilde{\Theta}_b^{-1} \sum_{m=0}^\infty (- \widetilde{\Theta}_s \widetilde{\Theta}_b^{-1})^m
\label{s_sigman}
\end{equation}
where $ \widetilde{\Theta}_b$ and $\widetilde{\Theta}_s$ are the body and soul parts of $\widetilde{\Theta}$,
respectively.
In order to understand this relation we now express
 $\widetilde{\Theta}' $ directly using the dual embedding map $\widetilde{\Phi}'$.
 From the relation \eqref{s_dual} we have
\begin{eqnarray}
\widetilde{\Phi}' = (\widetilde{\Phi}^t \widetilde{J}_0 )^{-1} \widetilde{K},
\label{dual_map}
\end{eqnarray}
and thus
\begin{eqnarray}
\widetilde{\Theta}' & =& {\widetilde{\Phi}}^{'t} \widetilde{J}_0 {\widetilde{\Phi}}^{'} \nonumber \\
  & = & \widetilde{K}^t (\widetilde{\Phi}^t \widetilde{J}_0^t
  \widetilde{\Phi})^{-1} \widetilde{K}.
\label{dual_stheta}
\end{eqnarray}
Since
\[
\widetilde{J}_0^t = \begin{pmatrix} J_0 & 0\\
0 & \hat{J}_0 \end{pmatrix}^t = \begin{pmatrix} -J_0 & 0\\
0 & \hat{J}_0 \end{pmatrix},
\]
we can write
$\widetilde{\Phi}^t \widetilde{J}_0^t   \widetilde{\Phi} = - \Theta_B + \Theta_F$.
Therefore
\eqref{dual_stheta} can be written as
\begin{eqnarray}
\widetilde{\Theta}'
  & = & \widetilde{K}^t (- \Theta_B + \Theta_F )^{-1} \widetilde{K} \nonumber \\
  & = & - \widetilde{K}^t \widehat{\Theta}_b^{-1} \sum_{m=0}^\infty
  (\widehat{\Theta}_s \widehat{\Theta}_b^{-1})^m  \widetilde{K}
  ,
\label{inv_dstheta}
\end{eqnarray}
where $\widehat{\Theta}_b$ is the body part of $\Theta_B$ and $\widehat{\Theta}_s$ is
the soul part of $ - \Theta_B + \Theta_F $.
Note that the body part of $\Theta_B$ and that of $\widetilde{\Theta}$ are the same.
%
 Since $\widehat{\Theta}_b^{-1}$ in \eqref{inv_dstheta} and $\widetilde{\Theta}_b^{-1}$
 in \eqref{s_sigman} are the same,
 $\widetilde{\Theta}^{-1}$ is just differ from $\widetilde{\Theta}'$
in the soul part up to
the action of  $\widetilde{K} \in GL(n,{\Z})$.
%
This shows that $\sigma_n$ generates a Morita equivalent torus.

In general, one can apply the bosonic T-dual transformations partially such as to
the coordinates $\hat{Y}^{i}, \ \ 1 \leq i \leq 2p$  with $2p<n$ as in \eqref{part-dual}.
We assume now $n=2p+q$. Then the noncommutativity  parameters $\Theta$ is given and transformed by
\eqref{Etrans2}, where $\Theta_a$ is  $2p \times 2p$ and $\Theta_d$ is $q\times q$.
In this case, we consider  $\sigma_{2p} \in SO(n,n,{\cal V}_{\Z}^0)$
given by
\[
\sigma_{2p} =\begin{pmatrix}
0& 0 & I_{2p} & 0\\ 0& I_q & 0 & 0\\ I_{2p} & 0 & 0 & 0\\
0& 0 & 0 & I_q\\
\end{pmatrix} .
\]
One can easily check that the action of $\sigma_{2p}$  defined by \eqref{gaction_theta} yields
$\Theta'$ in \eqref{Etrans2}; $\sigma_{2p} \Theta =\Theta'$.
Still we have to show that this transformed $\Theta'$ corresponds to a Morita equivalent torus.
As it was shown in \cite{rfsz98} (see also \cite{ekn10}), this transformed
 $\Theta'$ can be obtained from a dual embedding map $ \Phi'$ satisfying \eqref{b_dual};
\begin{equation}
\label{embedmap tpub}
 \Phi' = \begin{pmatrix}  J_0 (T_{a}^t)^{-1} & - J_0 (T_{a}^t)^{-1} T_{b}^t  \cr
    0  & I_q \cr  0 &   T_{c}^t \end{pmatrix} ,
\end{equation}
where $J_0$ is the $2p\times 2p$ matrix defined before, and
  $T_{a}$ is $2p\times 2p$, $T_{b}$ is $q\times 2p$,  $T_{c}$ is $q\times q$ such that
 $T_{a}^t J_0 T_{a} := - \Theta_{a}$,
 $T_{b} := \Theta_{b}^t$, and $\Theta_{d} := T_{c}^t - T_{c}$.
 Namely one can check that $(\Phi')^t J \Phi' = \Theta'$ where
 \[
J =\begin{pmatrix}
(J_0)_{2p}& 0 & 0\\ 0& 0 & I_q \\
0& -I_q  & 0 \\
\end{pmatrix} ,
\]
and
 $\Phi^t J \Phi = -\Theta$ when the original embedding map $\Phi$ is given by
 \begin{equation}
\label{embedmap}
 \Phi = \begin{pmatrix}  T_{a} & 0  \cr
    0  & I_q \cr  T_b &   T_{c} \end{pmatrix} .
\end{equation}
Therefore $\Theta' = \sigma_{2p} \Theta $ and $\Theta$ are Morita equivalent.
Since the  noncommutativity parameters with the same body parts yield
Morita equivalent tori, we can say that the general T-dual transformations given by
\eqref{Etrans2} and \eqref{Ftrans2} yield  Morita equivalent noncommutative supertori.

The fact that the same body parts up to elements in
${\cal V}^0$ yield equivalent tori dictates us  another  symmetry
action of the following element of $SO(n,n,{\cal V}_{\Z}^0)$
\[
\nu(\widetilde{N})=\begin{pmatrix}
I_n & \widetilde{N}\\
0 &I_n\\
\end{pmatrix} ,
\]
where $\widetilde{N}$
is an antisymmetric  $n \times n$ matrix whose entries are in ${\cal V}_{\Z}^0$.
 The action of $\nu(\widetilde{N})$ is given  by
\begin{equation}
 \nu(\widetilde{N}) \widetilde{\Theta} = \widetilde{\Theta} + \widetilde{N} .
\end{equation}
%


 Finally, we consider the ``rotation" $\rho(\tilde{R}) \in
SO(n,n,{\cal V}_{\Z}^0)$ given by
\[\rho(\tilde{R})=\begin{pmatrix}
\tilde{R}^t & 0\\
0 & \tilde{R}^{-1}\\
\end{pmatrix},
\]
where $\tilde{R} \in GL(n, {\cal V}_{\Z}^0)$.
%
%
It was shown in \cite{ekn10} that the  action of the
above element to
 a  basis $\{\vec{E}_i \}$ ($i=1,2, \cdots, n$) for
 a given torus with $ \widetilde{\Theta}$
yields an isomorphic  torus with
$\widetilde{\Theta}' = \tilde{R} \widetilde{\Theta} \tilde{R}^t $.

Thus we have shown that the
three elements of  $SO(n,n,{\cal V}_{\Z}^0)$, which are $\sigma_{n}$ (or $\sigma_{2p}$ with $2p\leq n$), $\rho (\tilde{R})$, and $ \nu
(\widetilde{N})$,
 yield Morita equivalent noncommutative $n$-supertori.
 Therefore, the bosonic and fermionic T-duality transformations that we considered in  section 2
 correspond to the Morita equivalence of noncommutative supertori related by the above symmetry
 transformations.
In the $n=2$ case, the above three elements generate the group $SO(2,2,{\cal V}_{\Z}^0)$.
\\

\section{Conclusion}

In this paper, we show that under the bosonic and fermionic T-duality transformations
the relation between the corresponding dual background fields can be dictated by
the Morita equivalence of noncommutative supertori that can be constructed
with the dual background fields.
Especially, when we restrict ourselves to the duality transformations along two coordinate directions
only, then we obtain the symmetry group $SO(2,2,{\cal V}_{\Z}^0)$ which is
the symmetry group of the Morita equivalence of noncommutative supertori in two dimensions.
However the problem of the boundary condition for D-branes still remains.
We have used the hybrid formalism in four dimensions
since it is easier to analyze than the
full ten-dimensional pure spinor formalism.
The above problem does not appear in the ten-dimensional formalism.


We have discussed the extended T-duality in tree level,
\textit{i.e.} in the worldsheet without genus.
In the case of bosonic T-duality, it is the exact symmetry for all order of
the genus expansion when the dualizing coordinate $X^{i}$ is compact.
The holonomy of the auxiliary gauge field
$\int_{C}A^{i}dz+\bar{A}^{i}d\bar{z}$
gives the winding number, where $C$ is the nontrivial cycle on
the worldsheet. For fermionic T-duality, in order to extend it to all genus,
one needs the non-periodic fermionic variable satisfying
$\theta\to\theta+\xi_{C}$ when $\theta$ goes around the cycle $C$
as discussed in \cite{bm08}.

If the superstring background becomes nonconstant, then the symmetry structure
of the bosonic and the fermionic T-duality might be different from what we have considered in
this paper. It would be interesting
to extend to a more general case with extended symmetry of T-duality as mentioned or
the supergroup duality considered in \cite{Fre:2009ki}.
A supersymmetric extension of non-abelian T-duality \cite{de la Ossa:1992vc} which is recently extended
to R-R background \cite{Sfetsos:2010uq} would be also interesting. \\
%




%
\begin{appendix}
\section*{Appendix: Boundary Condition for D-branes}
In this appendix, we
discuss the boundary condition of the worldsheet action \eqref{ws2}
when spacetime-filling D-branes are present. In particular we examine
how the surface term \eqref{st} changes the boundary condition. 
Since the surface term \eqref{st} consists
of the fermionic fields only,
we focus on the boundary condition of the fermionic fields%
\footnote{
The boundary condition for bosonic fields is the same as the case
of usual noncommutativity \cite{SeWi} except that $X^{\mu}$ is replaced
by $Y^{\mu}$.
}.
The relevant part of the action is
\begin{gather}
\frac{1}{2\pi\alpha'}\int d^{2}z\,\Bigl(
-q_{\alpha}\tilde{\partial}\theta^{\alpha}
-\tilde{q}_{\alpha}\partial\tilde{\theta}^{\alpha}
+2\pi\alpha' F^{\alpha\beta}q_{\alpha}\tilde{q}_{\beta}\Bigr).
\label{chiral}
\end{gather}
As it is studied in \cite{OoVa, Se, BeSe},
the fields $q_{\alpha}$ and $\tilde{q}_{\alpha}$ can be integrated out
using their equation of motion
\begin{gather}
\tilde{\partial}\theta^{\alpha}=2\pi\alpha' F^{\alpha\beta}\tilde{q}_{\beta},
\quad
\partial\tilde{\theta}^{\alpha}=-2\pi\alpha' F^{\alpha\beta}q_{\beta},
\label{cm}
\end{gather}
which leads to the following action
\begin{gather}
S_{\mathrm{eff}}=\frac{1}{(2\pi\alpha')^{2}}\int d^{2}z\,(F^{-1})_{\alpha\beta}
\partial\tilde{\theta}^{\alpha}\tilde{\partial}\theta^{\beta}.
\end{gather}
The equation of motion for $\theta^{\alpha}$ and $\tilde{\theta}^{\alpha}$
becomes $\partial\tilde{\partial}\theta^{\alpha}=
\partial\tilde{\partial}\tilde{\theta}^{\alpha}=0$.
Then the solution of $\theta^{\alpha}$ and $\tilde{\theta}^{\alpha}$
are written as the sum of the holomorphic part and the anti-holomorphic part
as
\begin{gather}
\theta^{\alpha}(z,\tilde{z})=
\theta^{\alpha}_{L}(z)+\theta^{\alpha}_{R}(\tilde{z}), \quad
\tilde{\theta}^{\alpha}(z,\tilde{z})=
\tilde{\theta}^{\alpha}_{L}(z)+\tilde{\theta}^{\alpha}_{R}(\tilde{z}).
\end{gather}

When we consider the case of open string attached on the D-brane,
the surface term in the equation of motion is
\begin{gather}
(F^{-1})_{\alpha\beta}\Bigl(
\tilde{\partial}\theta^{\alpha}\delta\tilde{\theta}^{\beta}
+\partial\tilde{\theta}^{\alpha}\delta\theta^{\beta}
\Bigr)\Bigr|_{z=\tilde{z}}=0.
\label{bt}
\end{gather}
The above is satisfied by choosing the boundary condition as 
\begin{align}
\theta^{\alpha}(z=\tilde{z})=\tilde{\theta}^{\alpha}(z=\tilde{z}), \quad
\tilde{\partial}\theta^{\alpha}(z=\tilde{z})=
-\partial\tilde{\theta}^{\alpha}(z=\tilde{z}),
\label{bc}
\end{align}
where the second condition comes from
$q_{\alpha}(z=\tilde{z})=\tilde{q}_{\alpha}(z=\tilde{z})$
to preserve the half of supersymmetry. Under the boundary condition
\eqref{bc}, the two point correlators are given in \cite{OoVa, Se,BeSe}.
Here we focus the one of them as
\begin{gather}
\langle\theta^{\alpha}_{L}(z)\tilde{\theta}^{\beta}_{R}(\tilde{w})\rangle
=
-\frac{(2\pi\alpha')^{2}F^{\alpha\beta}}{\pi i}\log(z-\tilde{w}).
\label{corr}
\end{gather}
The contribution to the boundary condition from the surface term
\eqref{st} is
\begin{gather}
\delta S_{\mathrm{b}}=
\frac{1}{(2\pi\alpha')^{2}}\int dx\, (f^{-1})_{\alpha\beta}\Bigl(
\partial\tilde{\theta}^{\alpha}\delta\theta^{\beta}
+\tilde{\partial}\theta^{\alpha}\delta\tilde{\theta}^{\beta}
+\partial\theta^{\alpha}\delta\tilde{\theta}^{\beta}
+\tilde{\partial}\tilde{\theta}^{\alpha}\delta\theta^{\beta}
\Bigr)\Bigr|_{z=\tilde{z}=x}.
\label{st2}
\end{gather}
The first and second terms have the same form with \eqref{bt}
and then cancel each other because of the boundary condition \eqref{bc}.
So if we can have
\begin{gather}
\partial\theta^{\alpha}(z=\tilde{z})=
-\tilde{\partial}\tilde{\theta}^{\alpha}(z=\tilde{z}),
\label{bc2}
\end{gather}
then the third and fourth terms also cancel each other,
the surface term \eqref{st} does not change
the boundary condition \eqref{bc}.
On the other hand if \eqref{bc2} holds, two point correlator \eqref{corr}
must vanish. This leads to contradiction.
However, as we mentioned in section 2,
if we start from the full ten-dimensional pure spinor formalism,
we do not need to add or remove the surface term since
the square of
the derivative of the fermionic coordinate $\theta$ is already present,
and thus the problem with the boundary condition disappears.
\end{appendix}

\vspace{5mm}
\noindent
{\Large \bf Acknowledgments}

\vspace{5mm} \noindent The authors thank  KIAS for hospitality
during the time that this work was done.
This work was supported by the National Research Foundation (NRF) of
  Korea grants funded by the Korean government (MEST) [
    NRF-2009-0075129 (E.\ C.-Y.),
NRF-2009-0084601 (H.\ N.), and NRF-2008-331-C00071 (H.\ S.) ].
%
\\




\end{document}